\documentclass[11pt,twoside]{article}  
\usepackage{adassconf}

\begin{document}   

%
%

\paperID{O11.2}

%
%
%
%

\title{Intelligent Information Retrieval}

%

\author{Michael J. Kurtz, Guenther Eichhorn, Alberto Accomazzi, Carolyn Grant, Edwin Henneken, Stephen S. Murray}
\affil{Harvard-Smithsonian Center for Astrophysics, 60 Garden Street, Cambridge, MA 02138}


\contact{Michael Kurtz}
\email{kurtz@cfa.harvard.edu }

%
%
%
%
%

\paindex{Kurtz, M. J.}
\aindex{Eichhorn, G. }     
\aindex{Accomazzi, A. }     
\aindex{Grant, C. }     
\aindex{Henneken, E. }     
\aindex{Murray, S. S. }

%
%

\authormark{Kurtz, et al.}




\begin{abstract}          

Since it was first announced at ADASS 2 the Smithsonian/NASA
Astrophysics System Abstract Service (ADS) has played a central role
in the information seeking behavior of astronomers.  Central to the
ability of the ADS to act as a search and discovery tool is its role
as metadata agregator.  Over the past 13 years the ADS has introduced
many new techniques to facilitate information retrieval, broadly
defined.  We discuss some of these developments; with particular
attention to how the ADS might interact with the virtual observatory,
and to the new myADS-arXiv customized open access virtual journal.

The ADS is at http://ads.harvard.edu

\end{abstract}


\section{Introduction}

Information is just noise if we do not know it exists, or how to use
it.  The Smithsonian/NASA Astrophysics Data System, in collaboration
with astronomy's data centers and publishers, has been working for
somewhat more than a decade towards making the information of
astronomy increasingly discoverable, available, and useful.

The test of an organization which provides its users with
information is whether it has provided the information which the users
need, and at what cost.  A cost benefit analysis of the information
system in astronomy, consisting of the ADS, the CDS, and the journals
(the data archives were not included in the analysis) shows a benefit
of about 7\%\ of the total cost of the entire discipline can be
ascribed to the joint effors of these organizations in creating the
current on-line system, compared with the paper system of a decade ago
(Kurtz, et al, 2005).  This ten times the total cost of running these
organizations, and about a factor of 100 times the differential cost
of making the services on-line.

Astronomy is currently creating another layer of on-line information
services, beginning with the archives, as opposed to the journals.
This effort, known as the ''Virtual Observatory,'' has likely much in
common with the efforts which preceeded it, and which now collaborate
with it.  The greatest lesson to be learned from the recent past is
that massive success is possible; however it is not inevitable.

\section{Information Concentration}

Concentrated, complex organized information is the basis of
intellegence.  Before an information system can provide sophisticated
tools to facilitate the search and retrieval of complex information,
it must actually have sufficient information about the information
(meta-data) to allow the information to be organized with a degree of
sophistication equal to the capabilities of the search.

From its inception the ADS has put a substantial fraction of its total
resources toward the task of collecting, creating, and organizing
meta-data.  Grant, et al (2000) list hundreds of sources of these
data; this is but a partial list.  Very often a single source can
require several years of patient negotiation with different entities
(e.g. the copyright holder and the publisher).

Collecting data from existing data providers consumes substantial
effort.  There is no widely used standard for meta-data transmission;
the OAI-PMH system (www.openarchives.org), while promising in concept,
actually is only used by a fraction ($\sim15\%\ $) of our sources, and
even with these each individual provider has unique extensions.  This
means that the responsibility for parsing the meta-data and putting it
into a standard format rests with the ADS staff.

This is different than for some other information services, where the
data sources are required to adhere to a standard format. Examples are
Citeseer (citeseer.ist.psu.edu) and Citebase (www.citebase.org).  The
difference resulting from the two approches is large.  ADS has been a
mature research tool since its inception; researchers have been able
to count on it to be up to date and complete for the past decade.  The
services which require provider compliance are still quite incomplete;
they function primarily as research tools for computer scientists
interested in text retrieval.a754-1.ps

This is also different than for Google Scholar (scholar.google.com)
which, like ADS, negotiates access to meta-data from providers, but
unlike ADS is not able to devote the resources to understanding, and
individually parsing, the peculiarities of each individual source.  For
the areas ADS covers its abilities far exceed those of Google Scholar.

There is a lesson here for other information services: Just do it!
Waiting for all the information to be included in the service to be
available in just the right form delays the full implementation of the
service by a substantial factor.  If a service is valuable enough to
offer, it is valuable enough for the service provider to make work.
This is not a totally overwhelming task; the three major data services
in astronomy today, ADS, CDS, and NED all function with relativly
small staffs (the ADS staff size is six), and all are able to put
substantial effort into data ingest.

\section{Information Retrieval}

Information retrieval is successful if you find the information you
need, not, as Mick Jagger might say, the information you want.  It is
also important that this be able to be done quickly, as the most
valuable commodity in any research endevour is the time of the best
researchers.
a754-1.ps
There are many different techniques for achieving this, some as simple
as making the send button on the web site a bright color.

The ADS has been developing and implementing these techniques, often
in advance of any other group, throughout its entire history, begining
with our use of combination of evidence (Belkin, et al 1995); partial
match, natural language techniques, astronomy specific ontologies, and
hyperlinks from the outset (Kurtz, et al 1993).  The connection with
CDS/SIMBAD (Grant, et al 1993, Wenger, et al 2000) was another early
achievement.

Linking directly to data has been a feature of the ADS since its
inception, with links to the {\it A\&A} data tables (Ochsenbein \&
Lequeux 1995) and to the archival data-sets in the ADIL (Plante, et al
1996) being early examples.

Much of the talk at the conference described the IR advances made by
ADS, the interested reader is referred to Kurtz, et al (2000) and
Kurtz, et al (2005) for historical summaries.

\section {$2^{nd}$ Order Operators}

Second order database operators (Kurtz, 1992) are database functions
which, instead of returning the data corresponding to a query,
performs some function on data associated with those data.  For
example a query might return a list of papers concerning redshift
surveys; these papers have lists of papers they reference.  A $2^{nd}$
order operator could take these lists, collate and sort them, then
return a list of those papers which are most cited by papers
concerning redshift suveys.  

Kurtz, et al (2002, 2005) discuss how using these operators within ADS
one can obtain the most recent, most popular, most useful, and most
educational papers on a complex topic of interest.

\begin{figure}
\epsscale{1.0}
\plotone{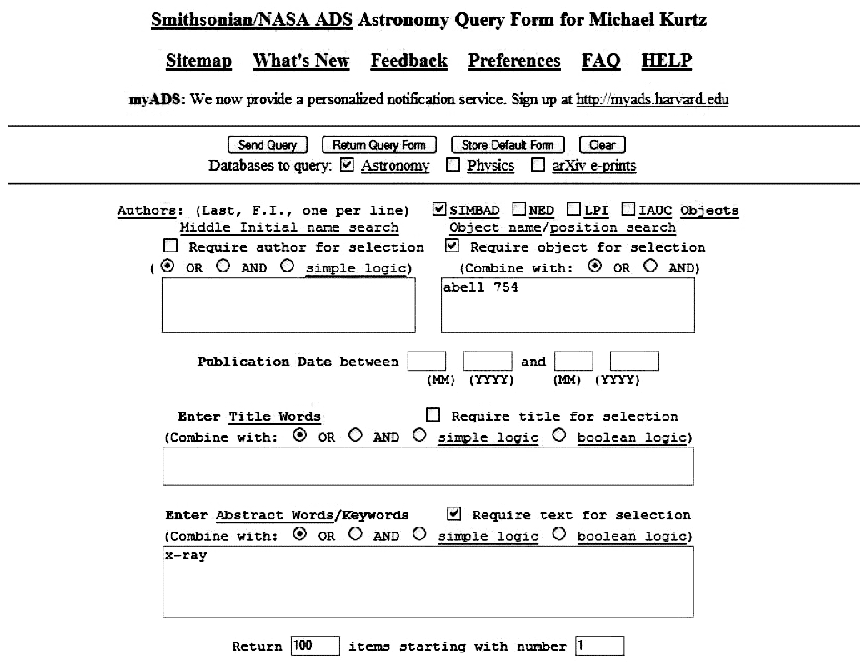}
\caption{A combined ADS-SIMBAD query asking for papers on the x-ray properties of Abell 754, with data links (not shown).}
\end{figure}

Here we demonstrate how to use ADS, and $2^{nd}$ order queries, to
discovery the most useful on-line data sets for a particular topic.
We begin, in figure 1, by asking for all articles with linked
data-sets (this requirement is not shown, it is futher down the form)
on the x-ray properties of the nearby cluster of galaxies Abell 754.
The result of this simple ADS/SIMBAD query is shown in figure 2, there
are 79 articles which meet this description.

\begin{figure}{b}
\epsscale{1.0}
\plotone{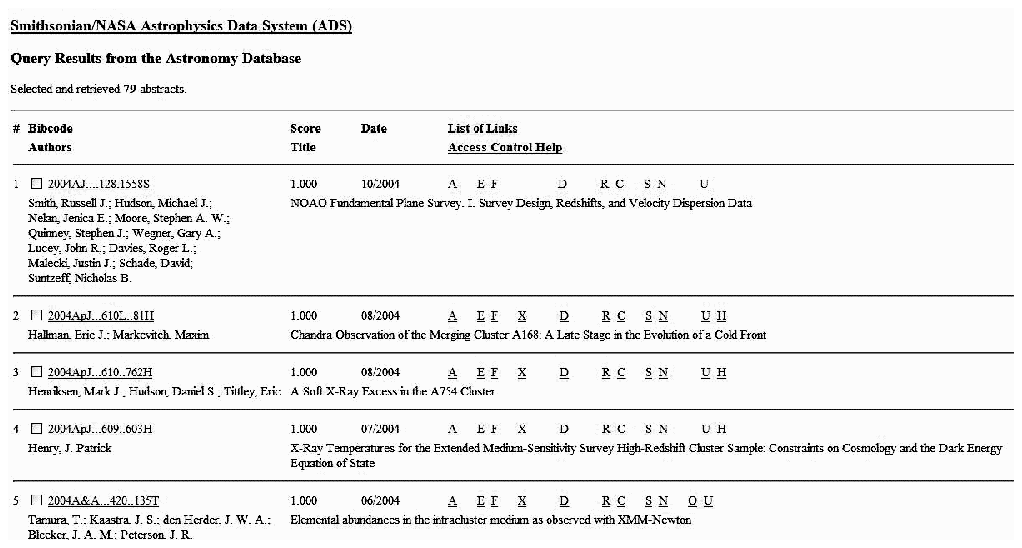}
\caption{The result of the A754/x-ray query, showing the most recent such papers}
\end{figure}

Now we are ready to use a $2^{nd}$a754-1.ps order operator.  We assert that, on
average, people who have read the articles in the list in figure 2 are
people interested in data about the x-ray properties of Abell 754. The
articles these people read the most, will, on average, be important to
us, as we are interested in data concerning the x-ray properties of
Abell 754.  Next we go to the bottom of that list, shown in figure 3,
we select all 79 articles from the list, and request the also-read
articles for that group.

\begin{figure}
\epsscale{1.0}
\plotone{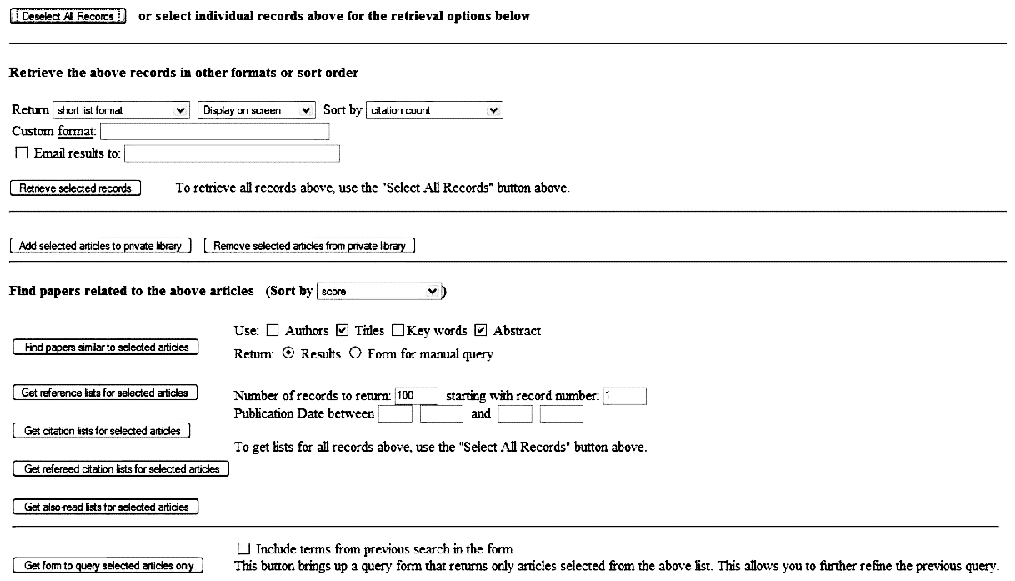}
\caption{The bottom of the results page, showing the $2^{nd}$ order queries}
\end{figure}

The result of that query is shown in figure 4.  The top five articles
are compendium articles on the x-ray properties of nearby clusters
from the Chandra, ROSAT, XMM, ASCA, and Einstein satellites, including
Abell 754.  This confirms our assertion above.  If we follow the ''D''
link for the top article, we go to the associated page in the Chandra
archive, shown in figure 5.  From here one can download a substantial
fraction of the Chandra data on nearby x-ray clusters, and one can
also access the software tools necessary to reduce and analyze it.

\begin{figure}
\epsscale{1.0}
\plotone{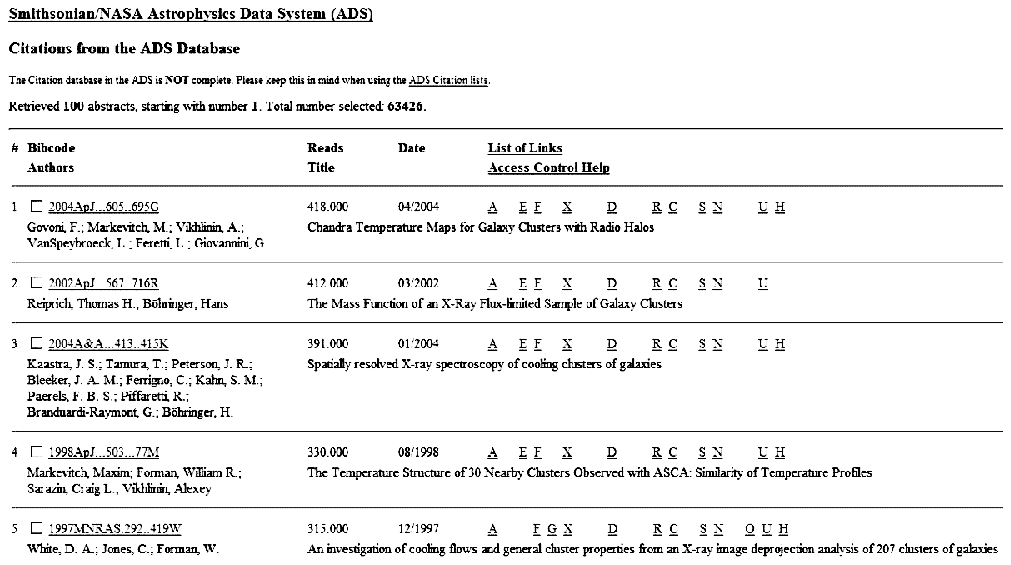}
\caption{The result of the $2^{nd}$ order query, compendium articles of data from the major recent x-ray telescopes on nearby clusters of galaxies, including Abell 754}
\end{figure}

The basic data used by this service is currently being created by the
archives, establishing links between data in the archives and
published papers.  Recently the Astrophysical Journal has established a
method of doing this directly by authors.

\begin{figure}
\epsscale{1.0}
\plotone{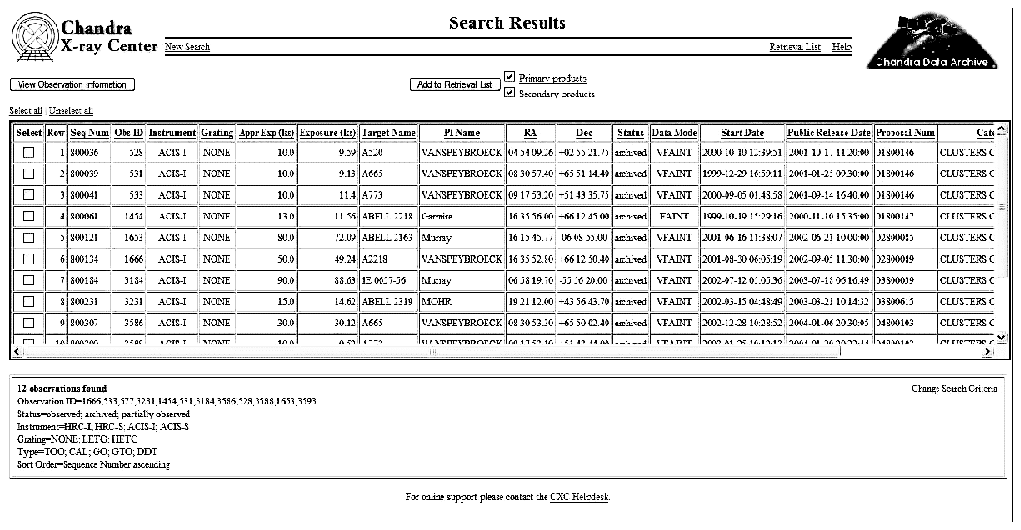}
\caption{The ''D'' link for the top article from the last query leads to the associated data page in the Chandra archive.}
\end{figure}

\section{myADS-arXiv}

Since late 2003 (Kurtz, et al 2003) the ADS has provided a
notification service, myADS, which uses the $2^{nd}$ order operators to
give our users a sophisticated and powerful means of keeping abreast
of the current literature in their sub fields of astronomy and
physics.

Beginning in the spring of 2005 we have extended this service with the
collaboration of the arXiv (www.arxiv.org) to include the arXiv
e-print service.  We now have three different, complimentary services:
myADS-Astronomy, myADS-Physics, and myADS-arXiv.

MyADS-arXiv is facilitated by a tight collaboration between ADS and
arXiv, this includes sharing of readership information, parsing the
reference lists, and merging the preprints with the journal articles
for the same article.  Once an article appears in the final published
source ADS adopts that version as definitive.

MyADS-arXiv is a fully customized (to each individual user), open
access virtual journal covering the most important papers of the past
week in physics and astronomy.  Between a fifth and a quarter of all
working astronomers already subscribe to myADS.  Figure 6 shows the
custom version belonging to the first author, for the week preceeding
the conference.  It is also possible to make a myADS page which is not
associated with an individual; for example the link
http://adsabs.harvard.edu/myADS/cache/278851069$\_$PRE.html points to
a myADS-arXiv page covering virtual observatory matters.  This can be
viewed as an automatically generated newsletter.

\begin{figure}
\epsscale{1.0})
\plotfiddle{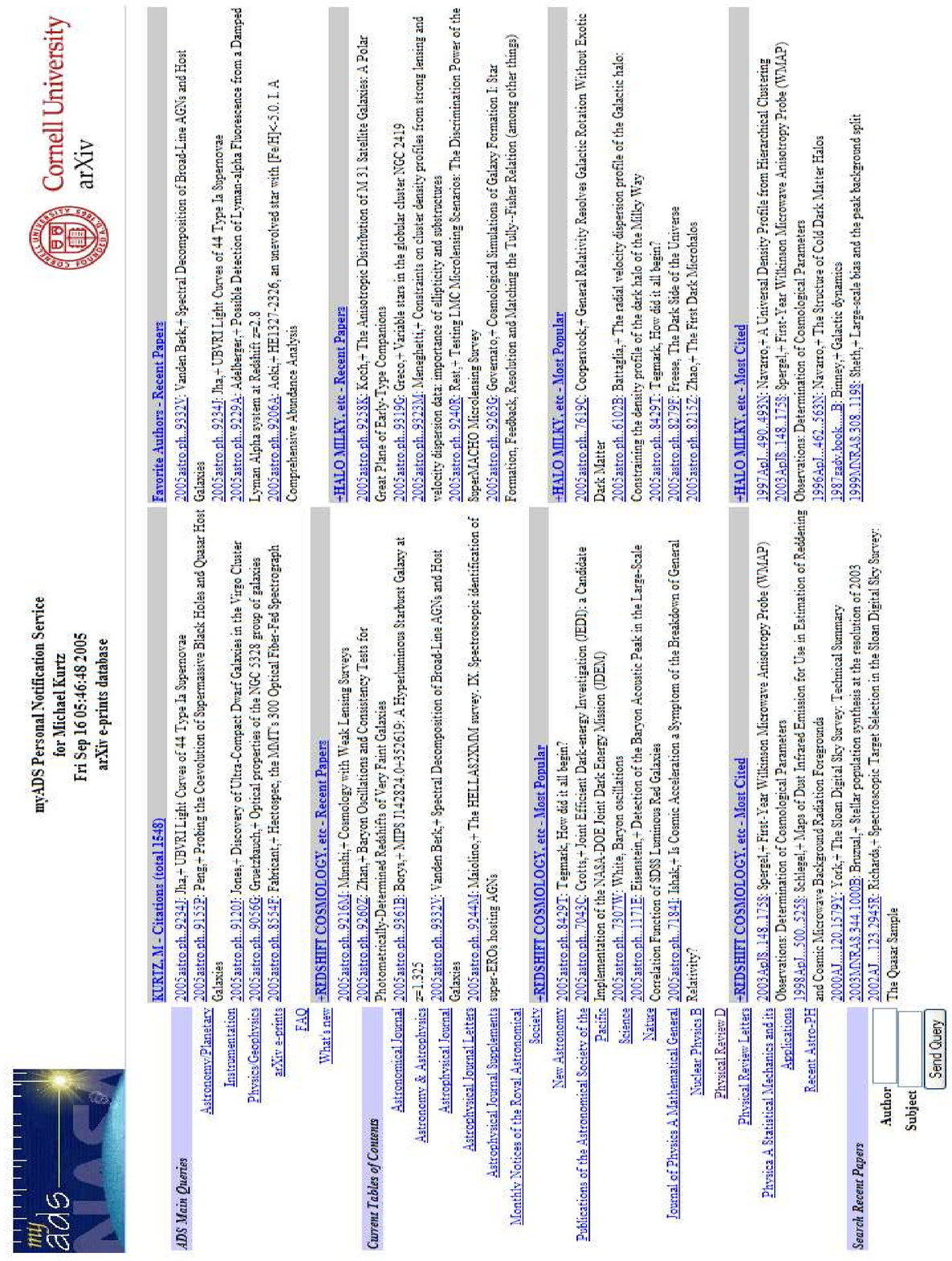}{5in}{270}{75}{85}{-360}{504}
\caption{An example of myADS-arXiv, for the first author of this paper} \label{myADS}
\end{figure}

\section{Discovery of Pointed Observations}

Szalay (2005) has presented the challenges and goals of a mode of
astronomy based on very large, homogeneous data-sets, such as the
SDSS, or in the future the LSST.  The scientific case for these
developments is exciting and compelling.

It seems likely, however, that for at least the next twenty years most
of the money and effort of astronomy will be devoted to pointed
observations with very large and expensive instruments.  Currently
HST, XMM, Keck, VLT, VLA and many other large pointeed instruments
dominate top level astronomy.  In the future, in addition to the mega
survey instruments, OWL, GMT, CELT, Con-X, JWST, ALMA and other very
costly pointed instruments will play a major role.

With the cost of dark time on a \$1,000,000,000 telescope running
around \$50,000 per hour the re-use of data in the telescope archives
would seem a highly cost effective means of doing research, as is now
the case with many current archives.

In order for astronomers to be able to find relevant data-sets from
pointed observations the ADS will index the abstracts of observing
proposals, and will point to the page in the archive devoted to that
project.  We have begun by indexing the proposals from several leading
archives.

We hope to do this in collaboration with with all interested groups,
and especially with the data centers and archives.  We ask for your
assistance and ideas on how to proceed.


\acknowledgments

The ADS is supported by NASA via NCC5-189 and by the Smithsonian
Institution.  MJK acknowledges partial financial support from the NVO
project.

\end{document}